\documentclass[12pt,preprint]{aastex}

\newcommand{\chandra}{{\it Chandra}}

\newcommand{\ee}[1]{\times10^{#1}}
\newcommand{\dem}{\mathcal{D}}

\newcommand{\kms}{km\,s$^{-1}$}

\newcommand{\thori}{$\theta^1$\,Ori~C}
\newcommand{\tnm}[1]{\tablenotemark{#1}}
\newcommand{\ud}[2]{$^{+#1}_{-#2}$}
\slugcomment{{\Large DRAFT: \today}}

\begin{document}

\title{Differential Emission Measure Determination of Collisionally
Ionized Plasma: II. Application to Hot Stars}

\shorttitle{DEM Determination: Hot Stars}

\author{Patrick S. Wojdowski\altaffilmark{1} and Norbert S. Schulz}

\affil{Center for Space Research, Massachusetts Institute of
Technology}

\email{pswoj@space.mit.edu}

\altaffiltext{1}{current address: Aret\'e Associates, P.O. Box 6024, Sherman Oaks, CA 91413} 

\begin{abstract}
In a previous paper we have described a technique to derive
constraints on the differential emission measure (DEM) distribution, a
measure of the temperature distribution, of collisionally ionized hot
plasmas from their X-ray emission line spectra.  We apply this
technique to the \chandra{}/HETG spectra of all of the nine hot stars
available to us at the time this project was initiated.  We find that
DEM distributions of six of the seven O stars in our sample are very
similar but that \thori{} has an X-ray spectrum characterized by
higher temperatures.  The DEM distributions of both of B stars in our
sample have lower magnitudes than those of the O stars and one,
$\tau$~Sco, is characterized by higher temperatures than the other,
$\beta$~Cru.  These results confirm previous work in which high
temperatures have been found for \thori{} and $\tau$~Sco and taken as
evidence for channeling of the wind in magnetic fields, the existence
of which are related to the stars' youth.  Our results demonstrate the
utility of our method for deriving temperature information for large
samples of X-ray emission line spectra.
\end{abstract}

\section{Introduction}
\label{sec:intro}
In hot stars, X-rays are emitted by gas heated in shocks in the wind.
In most cases, these shocks are thought to result from an instability
inherent in the mechanism which drives the wind: absorption of stellar
UV radiation in line transitions.  In some cases, however, other
physical processes may play a role in the heating of gas to X-ray
emitting temperatures.  For example, in very young hot stars,
channeling of the wind by magnetic fields may play a strong role in
heating the wind \citep[e.g.,][]{udd02}.  Among the evidence that
X-ray emission from some hot stars may be due to a mechanism other
than the instability of the radiative driving mechanism, is the fact
that the X-ray spectra of some hot stars are characteristic of
emission by plasma of higher temperature than others.

Line emission from hot plasmas is highly sensitive to temperature.
Therefore, insofar as theories of hot star winds make predictions for
the temperature of X-ray emitting plasmas in them, high-resolution
X-ray spectroscopy offers a powerful method with which to test those
theories.  However, there are two obstacles in using high-resolution
spectroscopy to study the winds of hot stars.  First, in most hot
stars, the winds from which the X-ray emission arises have velocities
in excess of 1000\,\kms.  Owing to this fact, X-ray emission lines are
broadened and blended, making it difficult or impossible to measure
fluxes of many emission lines \citep[e.g.,][]{cas01,kah01,sch03}.
Second, the theory of hot star winds is not sufficiently developed to
the point where detailed predictions for X-ray emission line spectra
can be made for direct comparison with observations.  

While there are open questions regarding the origin of the X-ray
emitting plasma in hot stars, it is well established that the emitting
plasma in hot stars is in collisional ionization equilibrium (CIE,
see, e.g., \citealt{pae03}).  In a previous paper \citep[][hereafter
Paper I]{woj04}, we have described a method for measuring the
differential emission measure (DEM) distribution, a measure of the
temperature distribution of a hot, optically thin plasma in
collisional ionization equilibrium (CIE).  In this method, a number of
templates, each of which contains all of the emission lines of a
single ion, are fit to an observed high-resolution spectrum.  The
best-fit normalization for each of these templates then provides a
measure of the product of the elemental abundance and a weighted
average of the DEM distribution over the range where the ion emits.
These measurements may then be plotted in a way that may be described
as a one-dimensional ``image'' of the DEM distribution.  This method
is well suited to the study of hot stars because the use of templates
makes use of the information available from line blends and because it
does not require that assumptions about the form of the DEM
distribution be made.

In \S\ref{sec:analysis} we describe our sample of nine early-type
stars which have been observed with \chandra/HETGS, the data, and the
application of the method of Paper~I to it.  In \S\ref{sec:discuss} we
discuss the implications of our results.

\section{Spectral Data Analysis}
\label{sec:analysis}

We obtained the data from the observations of all of the nine bright O
stars observed with the \chandra{} High-Energy Transmission Grating
Spectrometer (HETGS).  In Table~\ref{tab:data}, we list these nine
stars, the distances and interstellar absorption columns to them that
we adopt, and the effective exposure times, \chandra{} Observation ID
numbers, and any references for the \chandra/HETGS observations of
them.

For all of the stars in our sample, our analysis proceeds from binned
count spectra (PHA files), instrument effective areas and exposure
times (ARF files), and wavelength redistribution functions (RMF
files).  For \thori{} we use the spectral data and instrument
response files described by \citet{sch03}.  For all of the other
stars, we use the following procedure.  We use spectra extracted as
in the standard pipeline processing done by the \chandra{} X-ray
Center.  We calculate the appropriate effective area and exposure time
for the positive and negative first orders of the the High Energy
Grating (HEG) and the Medium Energy Grating (MEG) for each observation
using a standard procedure\footnote{see
\url{http://asc.harvard.edu/ciao/threads/}}.  For each of the two
gratings we add the spectra and effective areas for the positive and
negative first orders.  For those stars observed in two pointings, we
add the spectra and exposure times and average the effective areas for
the two pointings.  In all cases we use the standard redistribution
functions.  The background expected for these spectra is very low and
we make no correction for it.

\begin{deluxetable}{lrllrrrrrrrrr}
\rotate
\tablewidth{0pt}
\tabletypesize{\scriptsize}
\tablecaption{Stars and Data in our Study\label{tab:data}}
\tablehead{
& & & & \multicolumn{2}{c}{Adopted Values} & & & &
\multicolumn{4}{c}{References\tnm{k}} \\
\cline{5-6}\cline{9-12}
\colhead{HR \#} & \colhead{HD \#} & \colhead{Name} & \colhead{Type} &
\colhead{Distance} & \colhead{Column Density} & \colhead{Exposure} &
\colhead{Obs IDs} & \colhead{Type} & \colhead{Distance} & \colhead{Absorb} &
\colhead{Data\tnm{a}} \\
& & & & \colhead{(pc)} & \colhead{($10^{20}$\,cm$^{-2}$)} & \colhead{(sec)} }
\startdata
3165&  66811\phm{A} & $\zeta$ Pup      &O4I(n)f\tnm{f}&450&1.0&68,598&640&24(22)& 10,11 &15& 1 \\
1895&  37022\phm{A} & \thori{}\tnm{j}  &O4--6p var &450&19\phm{.0}&82,975& 3,4 &26(22)&15&5& 5,6 \\ 
8281\tnm{b} 
    & 206267A       & \nodata 	       &O6.5V((f))\tnm{d}&800&30\phm{.0}&72,557& 1888,1889&25(22)&12&16& 27 \\
2782&  57061\phm{A} & $\tau$ CMa       &O9II&1480& 5.8\tnm{c} &87,095&
2525,2526&24(22)&7&7& 27 \\
1899&  37043\phm{A} & $\iota$ Ori      &O9III\tnm{e}&440&2.0&49,917&  599,2420 &25(22)&13&15& \nodata \\
1852&  36486\phm{A} & $\delta$ Ori A   &O9.5II\tnm{h}&501&1.5&49,045&639&24(22)&15&15&4\\
1948&  37742\phm{A} & $\zeta$ Ori A    &O9.7Ib\tnm{g}&501&3.0&59,640&610&24(22)&15&15& 2 \\
6165& 149438\phm{A} & $\tau$ Sco       &B0.2V&132&2.7&59,630&638&23&  8(3) &15& 3 \\
4853& 111123\phm{A} & $\beta$ Cru      &B0.5III\tnm{i}&110&1.7&74,379&2575&18&  8(9) &14& \nodata \\
\enddata
\tablerefs{(1)~\citealt{cas01}; (2)~\citealt{wal01};
(3)~\citealt{coh03}; (4)~\citealt{mil02}; (5)~\citealt{sch00};
(6)~\citealt{sch01,sch03}; (7)~\citealt{moi01}; (8)~\citealt{per97};
(9)~\citealt{alc02}; (10)~\citealt{bra71}; (11)~\citealt{sch97};
(12)~\citealt{sim68}; (13)~consistent with \citealt{hil97} and
references therein; (14)~\citealt{cod76}; (15)~\citealt[BSC]{ber96};
(16)~\citealt{shu85}; 
(18)~\citealt{hil69};
(22)~\citealt{mai04}; (23)~\citealt{wal71}; (24)~\citealt{wal72};
(25)~\citealt{wal73}; (26)~\citealt{wal81}; (27)~\citealt{sch03b}
} 
\tablenotetext{a}{Previous publications describing these data from \chandra{}.}
\tablenotetext{b}{The designation HR 8281/HD 206267 includes four
stars which we resolve.  The spectrum is for HD 206267A.}
\tablenotetext{c}{derived using 
$N_{\rm H}=5.8\times10^{21}E_{B-V}$\,cm$^{-2}$}
\tablenotetext{d}{HD~206267A consists of stars of type O6.5V((f)) and
B0V in a 3.7 day orbit and a third component of type O8V (Harvin et
al., in prep.).} 
\tablenotetext{e}{spectroscopic binary}
\tablenotetext{f}{variable of BY Dra type}
\tablenotetext{g}{emission-line star}
\tablenotetext{h}{eclipsing binary}
\tablenotetext{i}{variable star of beta Cep type}
\tablenotetext{j}{a determination of type O7V for \thori{} has also
been made by \citet{con72}}
\tablenotetext{k}{References in parenthesis indicate a secondary
reference.}
\end{deluxetable}

\notetoeditor{table note i (variable star of beta Cep type) in
Table~\ref{tab:data} does not fit on the page.}

\begin{figure}
\plotone{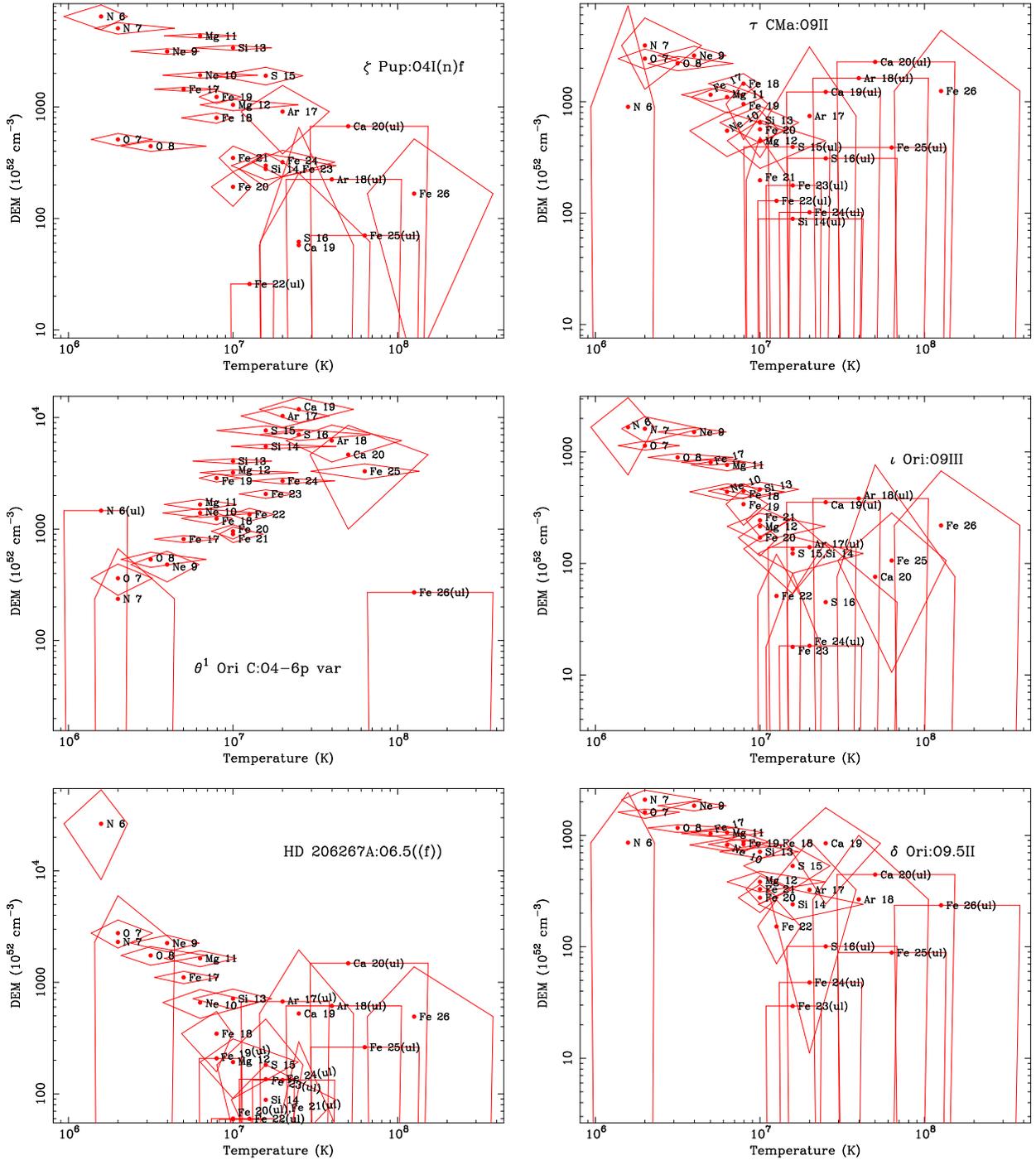}
\caption{Plots of $D$ as a function of temperature for $\zeta$ Pup,
\thori{}, HD~206267A, $\tau$~CMa, $\iota$~Ori, and $\tau$~Sco.  The
horizontal location of the data points is at the temperature where the
ion's emission peaks.  The horizontal extents of the diamonds give the
temperature range over which the ions' emission is significant and the
vertical extents of the diamonds give the statistical confidence
intervals for the values of $D$.  The designation ``(ul)'' indicates
an upper limit.}
\label{fig:dem_a}
\end{figure}

\begin{figure}
\includegraphics{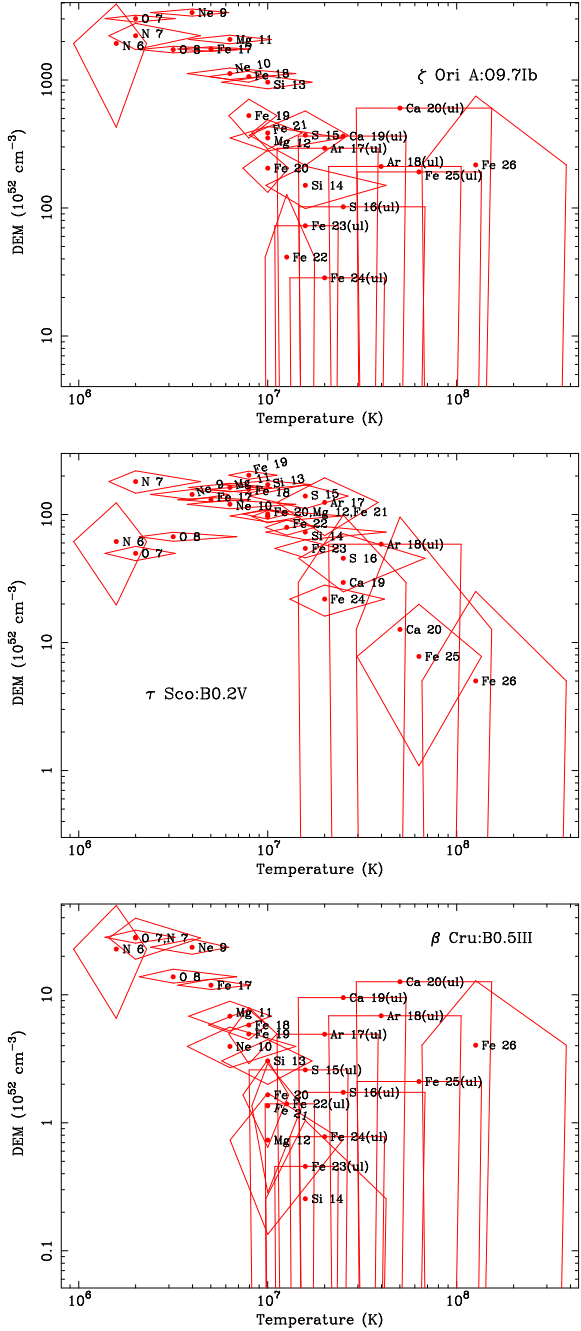}
\caption{DEM constraints for $\zeta$~Ori~A, $\delta$~Ori, and $\beta$~Cru.} 
\label{fig:dem_b}
\end{figure}

For each of the nine data sets, we apply the procedure described in
Paper~I --- i.e., we fit the spectra using templates, each of which
contains all of the emission lines of an ion.  In each of our fits to
the stellar data, we use two bremsstrahlung components except in the
case of $\tau$~Sco, \thori{}, and $\zeta$~Ori where three are
necessary to fit the continuum.  In each case, the model spectrum
includes neutral, interstellar absorption fixed at the value listed in
Table~\ref{tab:data}.  As explained in Paper I, the normalization
parameter of each emission line template obtained in the fit implies a
constraint on the product of the elemental abundance and a weighted
average of the DEM distribution:
\begin{equation}
D_{Z,z}\equiv{}A_Z\frac{\int{}_{-\infty}^\infty\theta_{Z,z}(T)\dem{}(T)d\log{}T}{\int_{-\infty}^\infty\theta_{Z,z}(T)d\log{}T}
\label{eqn:d_weight}
\end{equation}
where $Z$ is the atomic number of the element, $z$ is the charge
state, $A$ is the elemental abundance defined relative to solar,
$\dem{}$ is the differential emission measure distribution, $T$ is
temperature, and $\theta$ is the sum of the power functions (described
in Paper~I) for the lines of the ion.  In Figures~\ref{fig:dem_a} and
\ref{fig:dem_b} 
we plot $D$ vs. $T_{\rm p}$, where $T_{{\rm p},Z,z}$ is the
temperature where the function $\theta_{Z,z}$ peaks.  In all of
these plots, we use the same temperature range on the horizontal axis
as in the plots in Figures~2 and 3 of Paper I and, also as in those
figures, the vertical axes span three orders of magnitude.  Therefore
slopes of lines in these various plots may be compared.  As in Paper
I, we model the emission lines as Gaussians and allow the centers and
widths of all of the Gaussians to vary together.  The parameters which
describe the centers and widths of the lines are, respectively,
$v_{\rm r}$ and $v_{\rm t}$ and we tabulate the best-fit values of
these parameters in Table~\ref{tab:vtvr}.  The exact dependence of the
model line profiles on these parameters is given in equations~16 and
17 of Paper I.  While we have not quantitatively assessed the quality
of the fits, it may be seen from the spectral plots that the fits are
quite good considering the large number of lines apparent in the data.

As we have described in Paper~I, some excited states of some ions are
metastable and may undergo further excitation before decaying to
ground (pumping) and this leads the luminosities of some emission
lines to depend on physical variables other than the DEM distribution.
In the vicinity of a hot star, ions in many metastable states are
susceptible to excitation by the absorption of UV radiation from the
stellar photosphere.  As we have also described in Paper~I, we account
for any line pumping by letting the line ratios in the templates vary
as with collisional pumping with one free density parameter for each
ion.  This approach is somewhat ad hoc.  However, the fact that we
obtain good fits to the spectra indicates that pumping is well
accounted for.  Therefore, we do not expect significant errors in our
constraints due to line pumping or inaccuracies in our treatment of
it.

\begin{figure}
\plotone{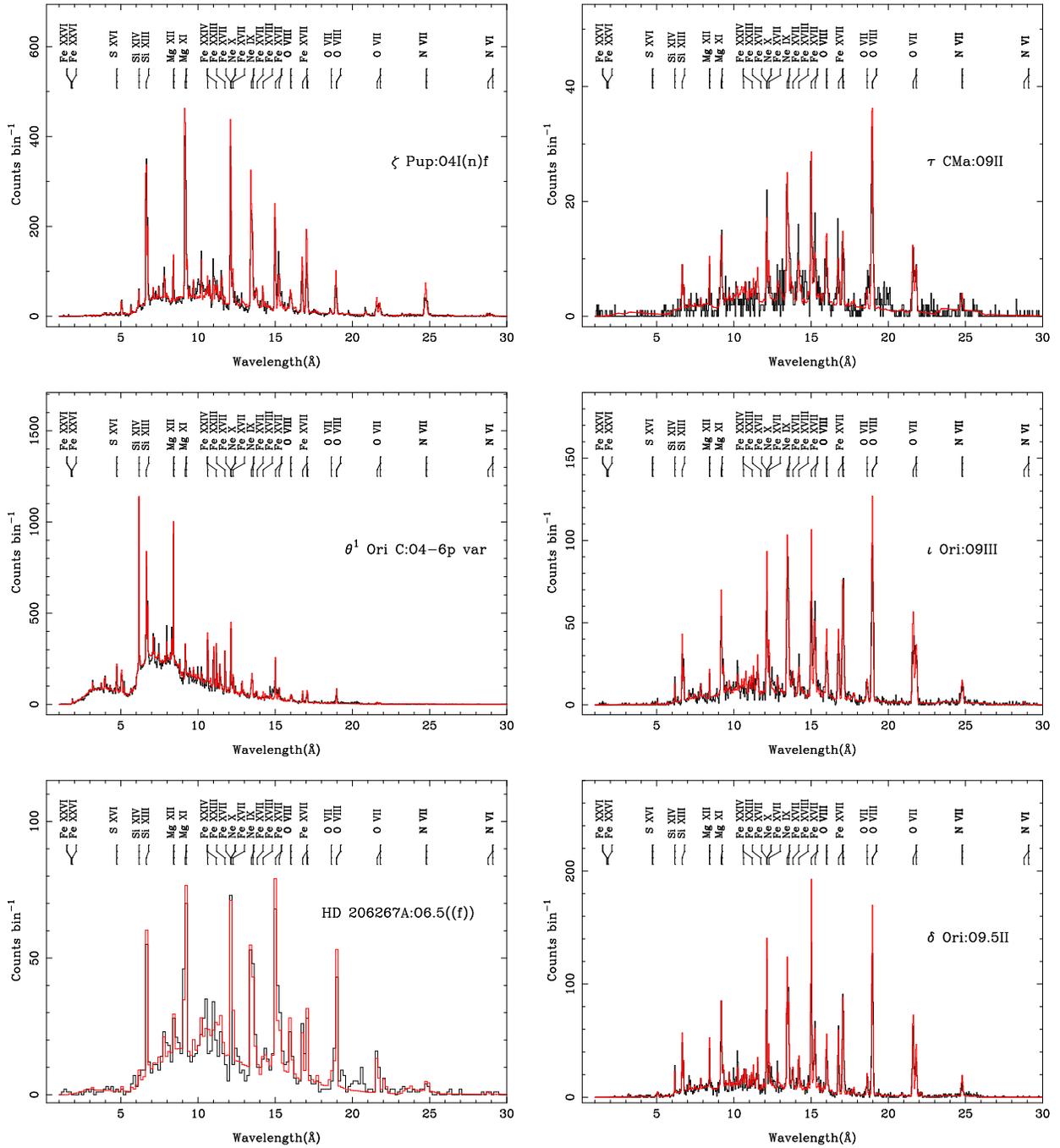}
\caption{MEG spectral data (black) and best-fit models (red) for
$\zeta$~Pup, \thori{}, HD~206267A, $\tau$~CMa, $\iota$~Ori, and
$\tau$~Sco. We include a number of line labels though not all of the
labeled lines are detected in every spectrum.}
\label{fig:spec_a}
\end{figure}
\begin{figure}
\includegraphics{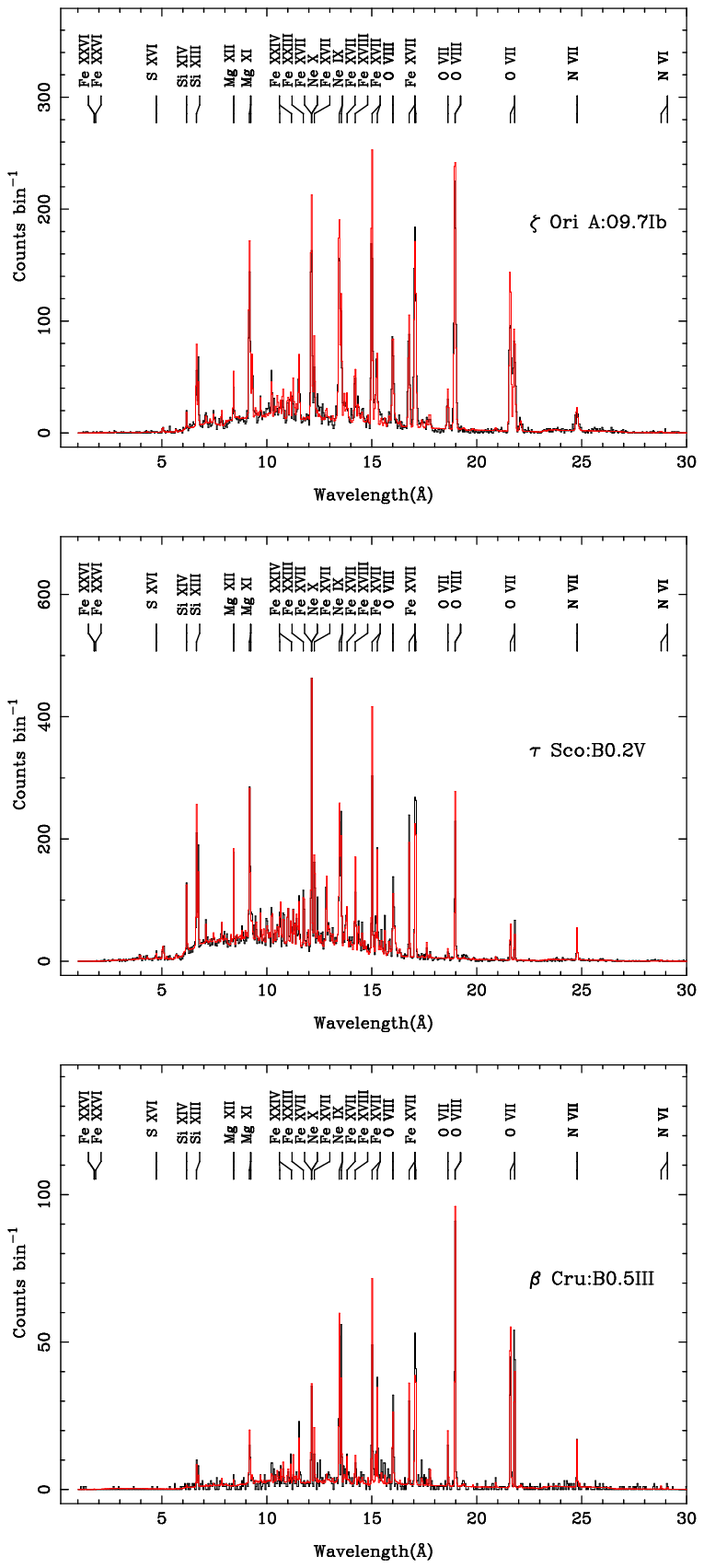}
\caption{MEG spectral data and best-fit models for
$\zeta$~Ori~A, $\delta$~Ori, and $\beta$~Cru.} 
\label{fig:spec_b}
\end{figure}

In the plots of $D$ vs. $T_{\rm p}$, for nearly every two ions with
close values of $T_{\rm p}$, the corresponding values of $D$ generally
appear to be consistent with each other or differ by no more than a
factor of two, indicating that the emission line spectrum is
consistent with emission from a plasma with solar abundances.  One
exception to this is the fact that, for $\zeta$~Pup, the values of $D$
for oxygen are approximately an order of magnitude less than the
values for ions with similar values of $T_{\rm p}$.  A similar result
has been obtained by \citet{kah01}.  The values of $D$ for nitrogen
appear to fit better with the rest of the data points than do the
values for oxygen.  Therefore, it is probably the case that oxygen is
underabundant in the wind of $\zeta$~Pup.  However, from this analysis
alone, it is impossible to exclude a hypothesis in which the DEM
distribution of $\zeta$~Pup peaks at approximately $3\ee6$\,K, the
abundance of nitrogen is greater than the solar value, and the
abundance of oxygen is near the solar value.  In addition, there is
significant scatter in the data points for $\zeta$~Pup in the
temperature range $4\ee6$--$10^7$\,K.  Some, but not all, of this
scatter might be accounted for by an underabundace of iron.  For
$\tau$~Sco, like $\zeta$~Pup, the values of $D$ for oxygen fall below
the values for other ions at nearby temperatures, though not by as
much.

Another exception to our results being consistent with solar abundance
is \thori.  For this star, most of the values of $D$ appear to lie on
a single curve except for those of the ions \ion{Fe}{20}--\ion{}{24}
which lie below it by a factor of approximately four.  \citet{sch03}
fit this spectrum using an explicit parameterized empirical model for
the DEM distribution with variable element abundances.  The best-fit
DEM distribution these authors derive has three peaks at 7.9, 25, and
63\,MK and troughs at 13 and 40\,MK and vanishes towards temperatures
higher and lower than the three peaks.  This three-peaked structure is
not immediately apparent in our constraints for \thori{} shown in
Figure~\ref{fig:dem_a}.  We have explicitly evaluated the right-hand
side of equation~\ref{eqn:d_weight} for the element abundances and DEM
distribution of \citet{sch03}\footnote{\citet{sch03}, in their
Figure~7 plot the ``DEM''.  However, this is not the DEM as we define
it (emission measure per temperature decade) but the emission measure
per temperature bin (with $\Delta\log{}T=0.05$).  Therefore, it must
be multiplied a factor of 20 to be compared with our results.}  and
while the values of $D$ for some ions have statistically significant
differences, there does not appear to be an overall systematic
difference between the values of $D$ obtained by the two methods.  In
particular, as with the values of $D$ obtained by our method, the
triply peaked DEM distribution is also not immediately apparent in the
values of $D$ obtained from the abundances and triply peaked DEM
distribution of \citet{sch03}.  This is somewhat puzzling.  However,
while a triply peaked DEM distribution is not immediately apparent in
our results, from a closer look it may be seen how a triply peaked DEM
distribution may be found with the method of \citet{sch03}.
In the method of \citet{sch03}, the formula for the quality of fit
includes a penalty for variations in the DEM distribution so, if
possible, elemental abundances will be adjusted to produce a flat DEM
distribution.  Variations in the DEM distribution will be found only
when lines of the same element require it.  Following our values of
$D$ for \ion{Fe}{17}--\ion{Fe}{24}, the first peak, the following
trough, and the rise to the second peak may be seen.  The trough after
the second peak may be seen in the drop from \ion{Ar}{17} to
\ion{Ar}{18} and from \ion{Ca}{19} to \ion{Ca}{20}.  The rise to the
third peak may be seen from the fact that the point for \ion{Fe}{25}
is higher than the one for \ion{Fe}{24}.

\begin{figure}
\includegraphics[angle=-90,width=3.0in]{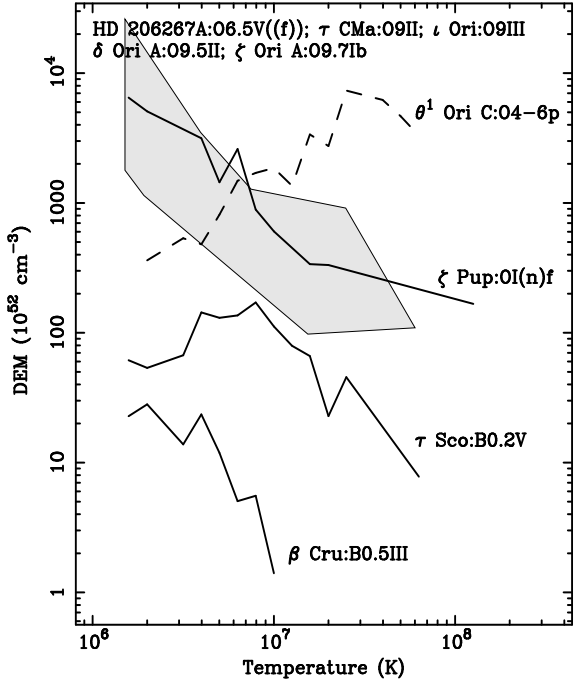}
\caption{A composite approximate representation of of the DEM
  constraints we derive for the nine stars.  The filled polygon
  encloses the lines for HD~206267A,
$\iota$~Ori, $\zeta$~Pup, $\zeta$~Ori~A, $\tau$~CMa, and
$\delta$~Ori~A which are very similar and difficult to show together.
The jagged shape of the curves gives some idea of the magnitude of the
  error for the DEM constraints.
Details of the construction of this figure are given in the text.}
\label{fig:comp_dems}
\end{figure}

In order to compare our constraints on the DEM distributions of the
stars in our sample with each other, we plot a composite of the nine
stars in Figure~\ref{fig:comp_dems}.  This figure is constructed as
follows.  For each star, we ignore those values of $D$ without lower
limits.  For the case of $\zeta$~Pup we also ignore the values of $D$
for oxygen.  Then, for those ions with the same value of $T_{\rm p}$
as given in Table~1 of Paper~I, we average the values of $D$ weighting
by the inverse square of the uncertainties.  The lines connect the
averaged values of $D$ for individual stars.  Because the method used
to construct this figure is ad hoc and because doing so would make the
figure harder to read, we do not attempt to include errors in this
figure.  However, the magnitude of the errors for this figure may be
inferred from Figures~\ref{fig:dem_a} and \ref{fig:dem_b} and may also
be inferred from the jagged curves.  The lines for HD~206267A,
$\iota$~Ori, $\zeta$~Pup, $\zeta$~Ori~A, $\tau$~CMa, and
$\delta$~Ori~A are very similar and difficult to show together.
Therefore, we show the line for $\zeta$~Pup and a filled polygon that
encloses the lines for the other five stars.  The axes and aspect
ratio of this plot are chosen so that slopes in this figure can be
compared with Figures 2 and 3 of Paper~I and Figures~\ref{fig:dem_a}
and \ref{fig:dem_b} of this paper.
With the exceptions of $\tau$~Sco and \thori, it may be inferred from
these plots that, in the temperature range $10^6$--$10^{8.5}$\,K, the
DEM distributions have peaks at temperatures of a few times $10^6$\,K
and decrease towards higher temperatures.  The DEM distribution of
$\beta$~Cru, in addition to having a smaller magnitude than the other
stars, declines more rapidly towards high temperatures.  For
$\tau$~Sco, the peak of the DEM distribution and the majority of the
emission appear to be at approximately $10^7$\,K.  Most of the
emission measure for \thori{} appears to lie near $3\ee7$\,K.  In
Figures~\ref{fig:spec_a} and \ref{fig:spec_b}, it may be seen that for
many of stars, the points for some or all of the ions \ion{Fe}{24},
\ion{Fe}{25}, and \ion{Fe}{26} are higher than those of lower charge
states of iron.  This suggests the possibility that these DEM
distributions increase at temperatures near $10^8$\,K.  However, all
of these high temperature points have low statistical significance and
it is not possible from this analysis to confirm the presence of any
such high temperature increase in the DEM distribution.

We give the best-fit line shifts and widths from our fits in
Table~\ref{tab:vtvr}.  If we add the systematic uncertainty of the
\chandra{} HETG absolute wavelength calibration (a few tens of
\kms{})\footnote{http://cxc.harvard.edu/proposer/POG/} to the
tabulated statistical confidence intervals, all of the line shifts we
measure are consistent with zero, with the exception of $\zeta$~Pup
and the possible exception of $\zeta$~Ori.  The blueshift we find for
$\zeta$~Pup is consistent with those reported by \citet{kah01} and
\citet{cas01}.  The latter authors attribute this blueshift to
obscuration of the far side of wind by photoelectric absorption in
nearer parts of the wind.
\begin{table}
\begin{tabular}{llrr}
\hline\hline
\multicolumn1c{Star} & \multicolumn1c{Type} & 
\multicolumn1c{$v_{\rm r}$ (\kms)} & \multicolumn1c{$v_{\rm t}$ (\kms)} \\
\hline
$\zeta$ Pup      & O5Iaf      & -460$\pm$20 & 770$\pm$20 \\
\thori{} & O6pe       & -29\ud{15}{14} & 326$\pm$18 \\
HD~206267A       & O6.5V((f)) & -10$\pm$140   & 1020\ud{140}{130} \\
$\zeta$ Ori A    & O9Iab      & -130$\pm$20 & 689$\pm$19 \\
$\tau$ CMa       & O9Ib	      & -120\ud{100}{110} & 990\ud{100}{90} \\
$\iota$ Ori      & O9III      & 60$\pm$40   & 930\ud{40}{30} \\
$\delta$ Ori A   & O9.5II     & 10$\pm$30   & 690$\pm$30 \\
$\tau$ Sco       & B0V        & 37$\pm$10   & 159\ud{14}{15} \\
$\beta$ Cru      & B0.5III    & -12$\pm$19  & 190$\pm$30 \\
\hline
\end{tabular}
\caption{Fit Line Shifts and Width Parameters}
\tablecomments{$v_{\rm r}$ and $v_{\rm t}$ are, respectively, the
  shift and Gaussian $\sigma$, see Paper~I.}
\label{tab:vtvr}
\end{table}
The blueshift we observe for $\zeta$~Ori~A may also be due to this
effect.  This possibility is discussed by \citet{wal01}.

\section{Discussion}
\label{sec:discuss}

We use a method, described in Paper~I, to determine the DEM
distribution for a sample of nine hot stars which works by fitting
entire X-ray spectra.  The fact that for all of the nine stars in our
sample good spectral fits are obtained suggests that the method is
free of significant systematic errors.  We find that the magnitudes of
the DEM distributions of the two B stars $\tau$~Sco and $\beta$ Cru
are much lower than those of the O stars as expected from previous
results \citep[e.g.,][]{ber96}.  For all of the stars in our sample
except for \thori{} and $\tau$~Sco we find the DEM distributions to be
peaked at a temperature of a few times $10^6$\,K or less, to decline
up to several times $10^7$\,K.  Though we cannot confirm it, the DEM
distributions of these stars may also increase at temperatures near
$10^8$\,K.  In contrast, we find that the peak of the DEM distribution
of $\tau$~Sco is near $10^7$\,K and the peak of the DEM distribution
of \thori{} is near $3\ee7$\,K.  For those two stars we also find that
the emission lines are narrow compared to stars of similar stellar
type.  High temperatures and narrow lines have previously been found
for $\tau$~Sco by \citet{coh03} and for \thori{} by
\citet{sch01,sch03} using these same data from \chandra.  In both
cases, it has been noted that the stars are young and that the high
temperatures and narrow lines have been attributed to channeling of
the wind in fossil magnetic fields.  In this scenario, wind travels up
the sides of magnetic loops, then shocks as wind from the two sides of
the loop collides.  Because the post-shock gas is stationary, narrow
lines are predicted.  The fact that the post-shock gas is stationary
also results in a larger velocity difference between the gas before
and after the shock and, therefore, higher temperatures \citep{udd02}.

The fact that the X-ray emitting plasma in the wind of $\tau$~Sco is
hotter than the X-ray emitting plasma in most stars has been apparent
from spectra of much lower resolution \citep*[e.g.,][]{coh97a,coh97b}.
A hot X-ray emitting plasma has also previously been detected from the
Orion Trapezium cluster which contains \thori{} \citep{yam96} though
it was not until a high-resolution image was obtained with \chandra{}
by \citet{sch01} that this hot plasma could be conclusively associated
with \thori.  Therefore, our finding that the temperatures of the
X-ray emitting plasmas \thori{} and $\tau$~Sco are high is neither a
new result of high-resolution spectroscopy nor of our analysis
technique.  However, as we have shown here, the method described in
Paper~I is useful in deriving well-defined quantitative constraints on
elemental abundances and differential emission measure distributions
from high-resolution X-ray spectra and, especially, large samples of
high-resolution X-ray spectra.  We expect that these results and the
results of similar observations and analyses will provide direction
for developing theory on X-ray emission in hot stars.

\acknowledgements

We thank John Houck for assistance implementing our analysis technique
in ISIS, Dan Dewey for a careful reading of the manuscript, David
Huenemoerder for providing data for the comparison of our results for
\thori{} with those of \cite{sch03} and the referee, Ehud Behar, for
helpful comments.  This research has made use of the National
Aeronautics and Space Administration's (NASA's) Astrophysics Data
System Bibliographic Services, the SIMBAD database, operated at CDS,
Strasbourg, France, and the Chandra Data Archive, part of the Chandra
X-Ray Observatory Science Center (CXC) which is operated for and on
behalf of NASA by the Smithsonian Astrophysical Observatory (SAO)
under contract NAS8-39073.  Support for this work was provided by NASA
through Chandra Award Number GO0-1119X by the CXC, through contract
NAS8-01129, and by the SAO contract SVI-61010 for the Chandra X-Ray
Center (CXC).

\bibliographystyle{apj} 
\bibliography{../ion_dem}

\end{document}